Thermoelectric Properties of Ho-doped Bi1-xSbx

K. C. Lukas, G. Joshi, K. Modic, Z. F. Ren, C. P. Opeil
Department of Physics, Boston College, Chestnut Hill, Massachusetts 02467
Los Alamos National Laboratory, Los Alamos, New Mexico 87545

Abstract

The Seebeck coefficients, electrical resistivities, total thermal conductivities, and magnetization are reported for temperatures between 5 and 350 K for n-type $Bi_{0.88}Sb_{0.12}$ nano-composite alloys made by Ho-doping at the 0, 1 and 3% atomic levels. The alloys were prepared using a dc hot-pressing method, and are shown to be single phase for both Ho contents with grain sizes on the average of 900 nm. We find the parent compound has a maximum of ZT = 0.28 at 231 K, while doping 1% Ho increases the maximum ZT to 0.31 at 221 K and the 3% doped sample suppresses the maximum ZT = 0.24 at a temperature of 260 K.

Introduction

Since the work of Smith and Wolfe [1], Bi-rich Bismuth Antimony n-type alloys have long been noted for their beneficial thermoelectric and magnetothermoelectric properties below room temperature [2]. Similar results were found for different Bi-Sb compositions of single crystal alloys prepared by different growth techniques [3,4]. New technological applications and work on $Bi_{1-x}Sb_x$ in zero field slowed for several years due to the lack of a comparable p-type material, because commercial devices rely on a combination of both n and p-type materials, whose overall thermoelectric figure of merit is given by [5]

$$ZT = \frac{(\alpha_p - \alpha_n)^2}{\left[(\rho_n \kappa_n)^{1/2} + (\rho_p \kappa_p)^{1/2}\right]^2}$$

in which α, ρ, and κ are the Seebeck coefficient, resistivity, and thermal conductivity, respectively. Interest in the alloy was revived by the discovery of high $T_c$ superconductors and the work of Dashevskii et al. showing that the total dimensionless figure of merit is approximately equal to the value of ZT for the n-type couple [6-9].

As group V metals, $Bi_{1-x}Sb_x$ forms a solid solution over the entire composition range, being semimetallic outside the range 0.07 < x < 0.22 where the alloy is semiconducting with the largest gap occurring around 17% Sb concentration [10-15]. The best thermoelectric properties are found for single crystals, with 0.09 < x < 0.16, when measured parallel to the trigonal axis [11,16,4]. Single crystals are not ideal for large scale manufacturing due to their difficulty in growing and cleaving, as well as their slow growth rate. More importantly, single crystals are mechanically weak [16], limiting their use in commercial applications. In an effort to increase their mechanical strength, polycrystalline alloys have been produced, but all have a lower ZT, dimensionless figure of merit, than single crystals. Several methods have been developed to improve ZT for n-type polycrystalline samples. These methods include: arc plasma [17], quenching [16], mechanical alloying [18], powder metallurgy [19], and doping [20,21]. Devaux et al. [17] studied the effects of grain size on the thermoelectric properties, specifically investigating the decrease of the lattice component of the thermal conductivity with the reduction of grain size. While the thermal conductivity was reduced due to phonon scattering, there was no benefit to ZT due to increased resistivity, as has also been seen in nanosized grains [22,23]. We present here the effects of doping atomic 1% and 3% Ho into nanopolycrystalline $Bi_{0.88}Sb_{0.12}$ prepared by ball milling and dc hot-pressing. The motivation for using Ho was to study the effects on the magnetothermoelectric properties due to a dopant with a large magnetic moment. Ho not only has a large magnetic moment, but it exhibits two magnetic transitions over the studied temperature range. Ho is antiferromagnetic over a large portion of the temperature range having a Neel temperature of approximately 133 K and is ferromagnetic at lower temperatures with a Curie temperature of about 20 K [24-28].

Experimental

Nano-polycrystalline $Bi_{1-x}Sb_x$ samples were prepared by ball milling and dc hot-pressing techniques described previously [29-32] where x=0.12 was determined to be the optimal Sb concentration for this process. Alloyed nanopowders were prepared by ball milling elemental chunks of bismuth (Bi) (99.99%, Alfa Aesar) and antimony (Sb) (99.99%, Alfa Aesar) with holmium (Ho) (99.9%, Alfa Aesar) according to the required composition $Bi_{0.88}Sb_{0.12}Ho_y$ (y = 0.01 and 0.03) for 5-12 hours, and then pressed at a temperature of $240^o$ C thereby creating disks 4 mm thick and 12 mm in diameter. X-ray diffraction was performed (Bruker AXS) in order to ensure the powders were alloyed into a single phase, and SEM (JEOL 7001F) images were taken of freshly fractured surfaces to observe the effects of grain growth during pressing. The pressed disks were then polished, chemically etched in a Bromine solution and metallic contacts were sputtered onto the faces.

From the disks, two samples were cut in order to measure the thermoelectric properties. Thermoelectric properties $\alpha$, $\rho$, and $\kappa$ were measured, in the standard two probe method, from 5-350 K using the thermal transport option of the Physical Properties Measurement System (PPMS) from Quantum Design. Samples for the thermal transport option were cut and measured perpendicular to the face of the disk (parallel with the press direction) with typical dimensions 3x3x4 mm. Gold coated OFHC (oxygen free high conductivity) copper disks provided by Quantum Design were soldered to the sputtered metallic contacts on the sample using Sn-Pb solder. A second sample was used to determine the Hall coefficient. The Hall coefficient was also determined using the PPMS, under a magnetic field of 9 T and a current of 20 mA. Our Hall sample was rotated by $180^o$ in field using the QD-PPMS AC Rotator option, thereby allowing the current direction to be switched and thus averaging out any anomalous effects on the measurement due to the field. These samples were prepared in a five wire Hall configuration with typical dimensions 1x3x11 mm. These samples were cut and measured perpendicular to the press direction. The carrier concentration was determined directly from the Hall coefficient. Resistivity, $\rho$ was measured using a standard four point probe technique, with the same sample and orientation used to attain the Hall coefficient, and mobility, $\mu$, was calculated from $R_H/\rho$. We assume the mobility to be essentially isotropic because the thermoelectric properties for alloys prepared by single dc hot-pressing techniques have been shown to be isotropic to within 10% [32]. Susceptibility was measured using the Vibrating Sample Magnetometer (VSM) option in the PPMS in a field of 0.1 T.

Results and Discussion

Figures 1 and 2 show the X-ray diffraction patterns and SEM images of $Bi_{0.88}Sb_{0.12}$, $Bi_{0.88}Sb_{0.12}Ho_{0.01}$ and $Bi_{0.88}Sb_{0.12}Ho_{0.03}$, respectively. X-ray peak positions confirm that the samples are alloyed and no second phases exist, while SEM images confirm that the average grain growth is proportional in each sample, and independent of the doping concentration. We confirm that the addition of Ho does not affect grain growth, which is known to alter thermoelectric properties [17]. The changes in $\alpha$, $\rho$, $\kappa$ are due to contributions from the magnetic moment, as well as the different size and mass of Ho. It should be noted that the grains are larger than usual for this process because the dc hot-pressing process was not optimized.

The data in Figure 3 shows the change in susceptibility due to the addition of Ho. Due to the large magnetic moment of Ho the susceptibility increases with the addition of 1% Ho and further increases with 3% Ho, where the Neel and Curie temperatures can also be seen more distinctly as the percentage of Ho increases. High enough temperatures could not be reached in the VSM to perform a Curie-Weiss fit as can be seen in the left inset of Figure 3. The top right inset plots Δcapacitance versus temperature for 1% Ho on a home built torque cantilever magnetometer, giving the same trend as the VSM.

The Hall coefficient is negative throughout the temperature range, showing that electrons are the majority carriers. The carrier concentration is presented in Figure 4. It can be seen that there is almost no change from the addition of 1% Ho at temperatures above 200 K while there is a decrease in carrier concentration below 200 K. The addition of 3% Ho increases the carrier concentration throughout the temperature range. The values for $Bi_{0.88}Sb_{0.12}$ are slightly higher than those reported in the literature for polycrystalline samples prepared by quenching and annealing [16], and three orders of magnitude higher than values reported for single crystals at 4.2 K [33]. Figure 5 shows the mobility, where again 1 and 0% Ho are nearly equal for higher temperatures (T>200 K) but deviate from one another at lower temperatures, where 0% Ho has a higher mobility than the 1% Ho sample. Again the 3% Ho doped sample has a consistently lower mobility throughout the temperature range. The values for mobility, Figure 5, are again lower than in the literature [16,33]; however, the $\log\mu_H$ vs. $\log T$ curve is linear for $Bi_{0.88}Sb_{0.12}$ showing the same qualitative trend throughout the temperature range. The addition of Ho causes a more defined change in the slope of the lines at 150 K, creating two different relationships for $\mu(T)$ above and below 150 K.

Figure 6 shows the temperature dependence of the resistivity for the different doping concentrations. Each sample shows typical semiconducting behaviour at lower temperatures, with the maximum value for $\rho$ at temperatures below 150 K increasing with doping concentration. At temperatures exceeding 150 K, the 0 and 1% Ho samples have nearly the same resistivity, while the 3% Ho sample always has a greater resistivity than the other compositions. The increase in resistivity at lower temperatures for the 1% Ho sample is due to both a decrease in carrier concentration as well as a decrease in mobility, while the increase in $\rho$ for the 3% Ho sample is due entirely to a decrease in mobility. The values for $\rho$ of $Bi_{0.88}Sb_{0.12}$ are higher than those of single crystal [11], which is typically the case, but comparable with those of different polycrystalline samples [16,17,19], including those from the work by Devaux et al. [20] and Dutta et al. [23].

The temperature dependence of the Seebeck coefficients are shown in Figure 7. Values for thermopower are negative over the entire range, confirming the majority carriers are electrons. The maximum values for the 0 and 3% Ho doping are approximately 125 µV/K at 150 K, while the maximum for 1% Ho is 145 µV/K at 100 K. The increase in the magnitude of the thermopower for 1% Ho is most likely caused by a decrease in carrier concentration at low temperatures, even though the mobility is not increased in this range. For 3% Ho, the carrier concentration is higher and the mobility is lower which should decrease the absolute value of α. Since α is nearly unchanged, it is apparent that phonon drag has a direct impact on the value of the Seebeck coefficient which can be seen in the increase in ρ due to phonon-electron interactions. The temperature at which there is a maximum for α in BiSb is dependent upon the concentration of Sb as well as ρ [11]. The absolute values for the maxima presented in Figure 7 are consistent with the literature values for both polycrystal [16,19,35] and single crystal samples measured perpendicular to the trigonal axis [11], although the maxima occur at a slightly higher temperature.

The temperature dependence of the thermal conductivity is shown in Figure 8 for all doping concentrations. We note that the thermal conductivity is drastically lowered, by more than four times, from that of single crystals [1,11] and is lower from other reported polycrystals [23,35]. Nanopolycrystalline samples prepared by cold-pressing rather than hot-pressing have a greater reduction in thermal conductivity, however ρ is drastically increased [22]. This is a product of the ball milling and dc hot-pressing preparation methods, where the size of the particles is very small which increases phonon scattering as described by Goldsmid et al. [36]. This is the case for both the 0 and 1% Ho samples. The thermal conductivity is further reduced in the 3% Ho sample where the lack of a peak at lower temperatures is evidence of structural disorder. The addition of 3% Ho to the sample lowers the phonon mean free path by being a point defect scattering mechanism, which has also been seen in the literature [1,4,35-37]. Ho is perhaps a more effective scatterer since Ho (164 amu) has an atomic mass in between that of Bi (208 amu) and Sb (121 amu). While there is a lowering in the thermal conductivity due to additional Ho ion scatterers, which has also been seen in other doped samples [35], the decrease in resistivity leads to an overall decrease in ZT.

Figure 9 gives the value for the dimensionless figure of merit. There is an increase in the thermoelectric figure of merit due to 1% Ho. The increase in ZT for the 1% sample at 200 K is due to the enhanced Seebeck coefficient while maintaining a similar value for ρ and κ as that of $Bi_{0.88}Sb_{0.12}$ for temperatures greater than 150 K. In contrast, ZT decreases for 3% Ho. Even though the thermal conductivity is decreased due to a shorter phonon mean free path, ZT is still lower due to additional point defect scatterers as well as phonon-electron interactions that affect the mobility of the charge carriers, increasing resistivity similar to results previously reported [17,22]. The 1% Ho sample presented here could be used with the p-type element of Hor et al. [35]; the benefits not only include improvement in the maximum value for ZT, by about 20\%, but also the temperature is the same at which both maxima occur, which is the ideal scenario of an optimized thermoelectric cooler or generator.

Conclusion

The effects on the thermoelectric properties of $Bi_{0.88}Sb_{0.12}$ due to the addition of Ho impurities have been studied and an improvement to ZT is found by the addition of 1% Ho. While improvements are found in thermoelectric properties in zero field, further changes may be found with the addition of magnetic field both parallel and perpendicular to the sample. Experiments on these samples, which will include the response of magnetic field on the unique magnetic properties of Ho, are underway.

Acknowledgements: The authors gratefully acknowledge M. S. Dresselhaus, J. C. Lashley and P. S. Riseborough for their fruitful discussions and careful reading of the manuscript as well as G.

McMahon for his assistance. This work is funded by the Air Force MURI program under contract FA9550-10-1-0533.)

List of Captions

Figure 1: XRD pattern for $Bi_{0.88}Sb_{0.12}$ (Black), $Bi_{0.88}Sb_{0.12}Ho_{0.01}$ (Red), $Bi_{0.88}Sb_{0.12}Ho_{0.03}$ (Green).

Figure 2: SEM images for (a) $Bi_{0.88}Sb_{0.12}$, (b) $Bi_{0.88}Sb_{0.12}Ho_{0.01}$ and (c) $Bi_{0.88}Sb_{0.12}Ho_{0.03}$. The scale bar is 1 μm.

Figure 3: Temperature dependence of the magnetic susceptibility for $Bi_{0.88}Sb_{0.12}$ (Black), $Bi_{0.88}Sb_{0.12}Ho_{0.01}$ (Red), $Bi_{0.88}Sb_{0.12}Ho_{0.03}$ (Green) given by the VSM option in the PPMS. The bump in the data at 10K is caused by the PPMS changing cooling modes and not by the sample. The upper right inset shows Δ capacitance vs. temperature on a home built torque cantilever magnetometer for $Bi_{0.88}Sb_{0.12}$ (Black) and $Bi_{0.88}Sb_{0.12}Ho_{0.01}$ (Red) giving the same trend as the VSM data. The left inset plots $1/\chi$ vs. temperature and shows that a Curie-Weiss analysis is not possible for this temperature range. The lower right inset gives an expanded view of $\chi$ vs. temperature from 150-300 K for $Bi_{0.88}Sb_{0.12}$ (Black) and $Bi_{0.88}Sb_{0.12}Ho_{0.01}$ (Red).

Figure 4: Temperature dependence of the carrier concentration of $Bi_{.88}Sb_{.12}$ (Black), $Bi_{0.88}Sb_{0.12}Ho_{0.01}$ (Red), $Bi_{0.88}Sb_{0.12}Ho_{0.03}$ (Green).

Figure 5: Temperature dependence of mobility for $Bi_{.88}Sb_{.12}$ (Black), $Bi_{.88}Sb_{.12}Ho_{.01}$ (Red), $Bi_{.88}Sb_{.12}Ho_{.03}$ (Green).
\linebreak

Figure 6: Temperature dependence of resistivity for $Bi_{.88}Sb_{.12}$(Black), $Bi_{.88}Sb_{.12}Ho_{.01}$ (Red), $Bi_{.88}Sb_{.12}Ho_{.03}$ (Green).

Figure 7: Temperature dependence of thermopower for $Bi_{0.88}Sb_{0.12}$(Black), $Bi_{0.88}Sb_{0.12}Ho_{0.01}$ (Red), $Bi_{0.88}Sb_{0.12}Ho_{0.03}$(Green).

Figure 8: Temperature dependence of thermal conductivity for $Bi_{0.88}Sb_{0.12}$(Black), $Bi_{0.88}Sb_{0.12}Ho_{0.01}$(Red), $Bi_{0.88}Sb_{0.12}Ho_{0.03}$(Green).

Figure 9: Temperature dependence of ZT for $Bi_{0.88}Sb_{0.12}$ (Black), $Bi_{0.88}Sb_{0.12}Ho_{0.01}$(Red), $Bi_{0.88}Sb_{0.12}Ho_{0.03}$(Green).

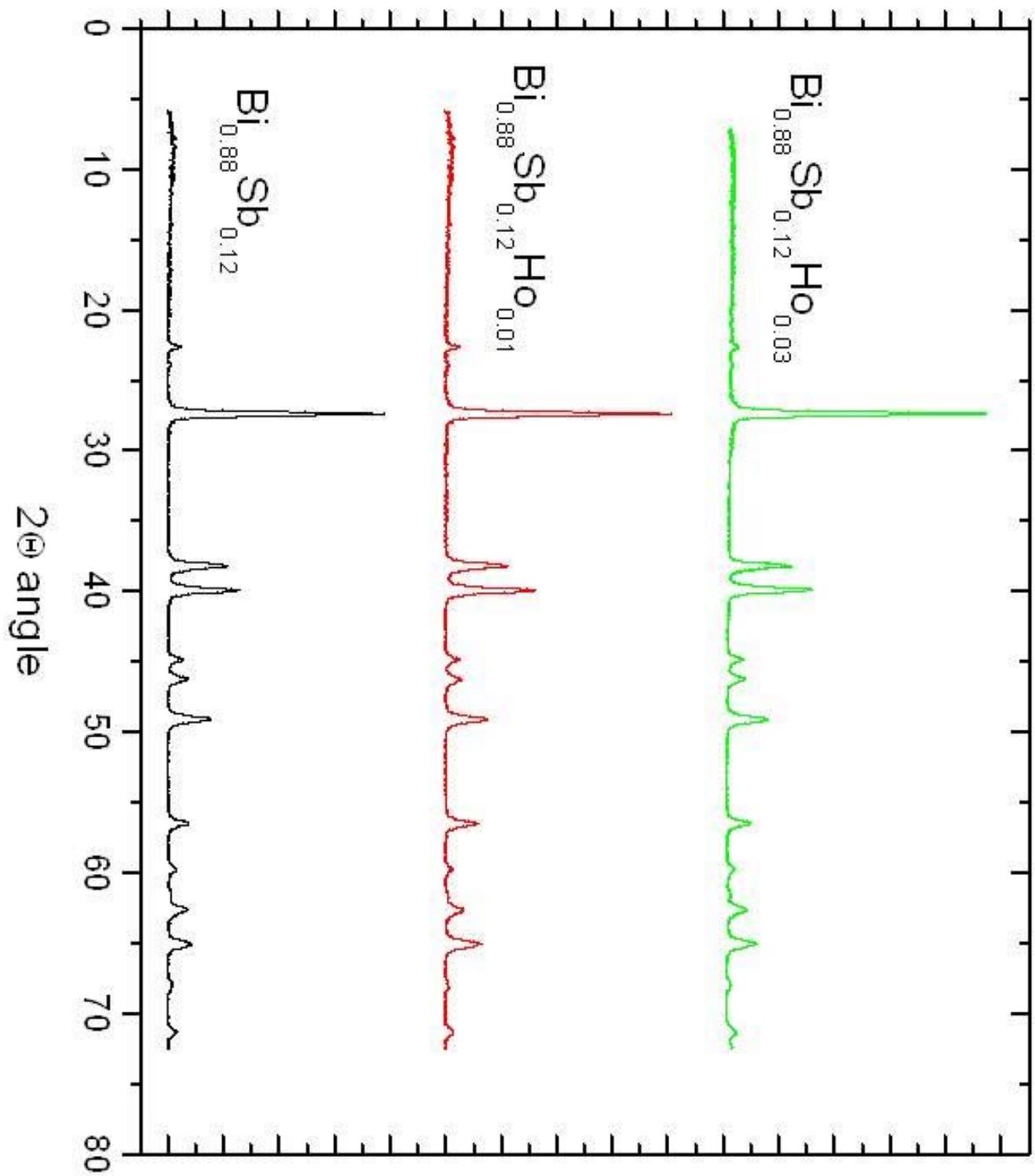

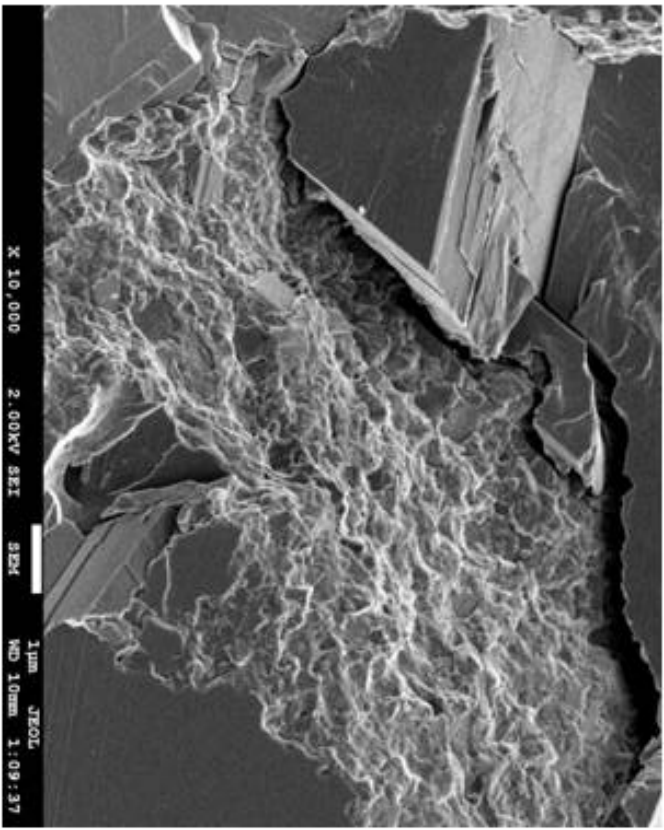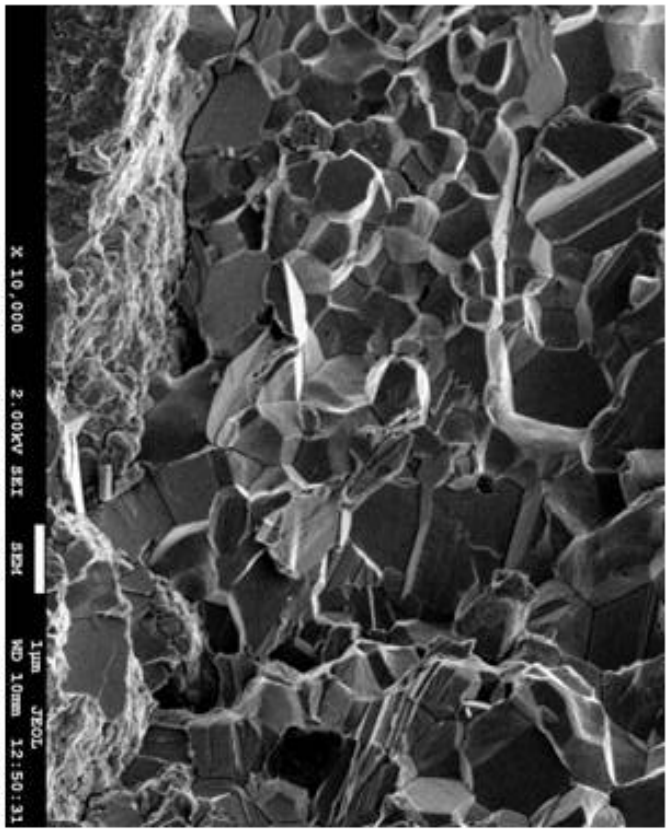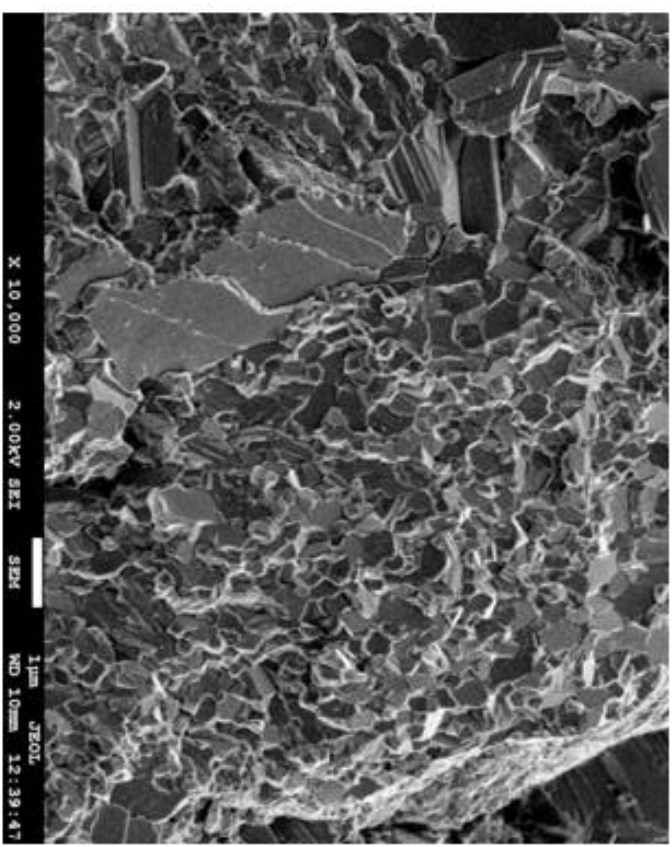

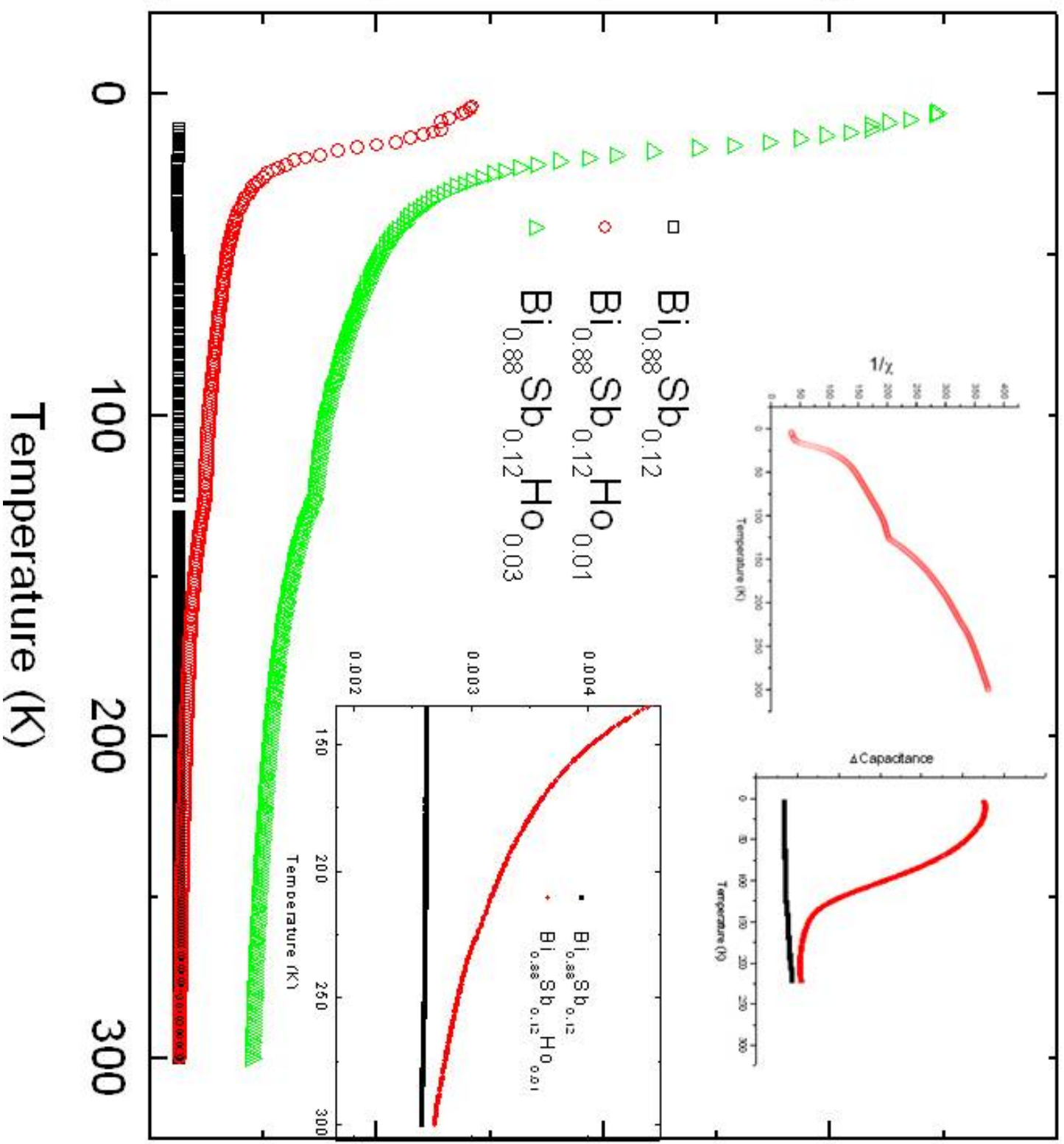

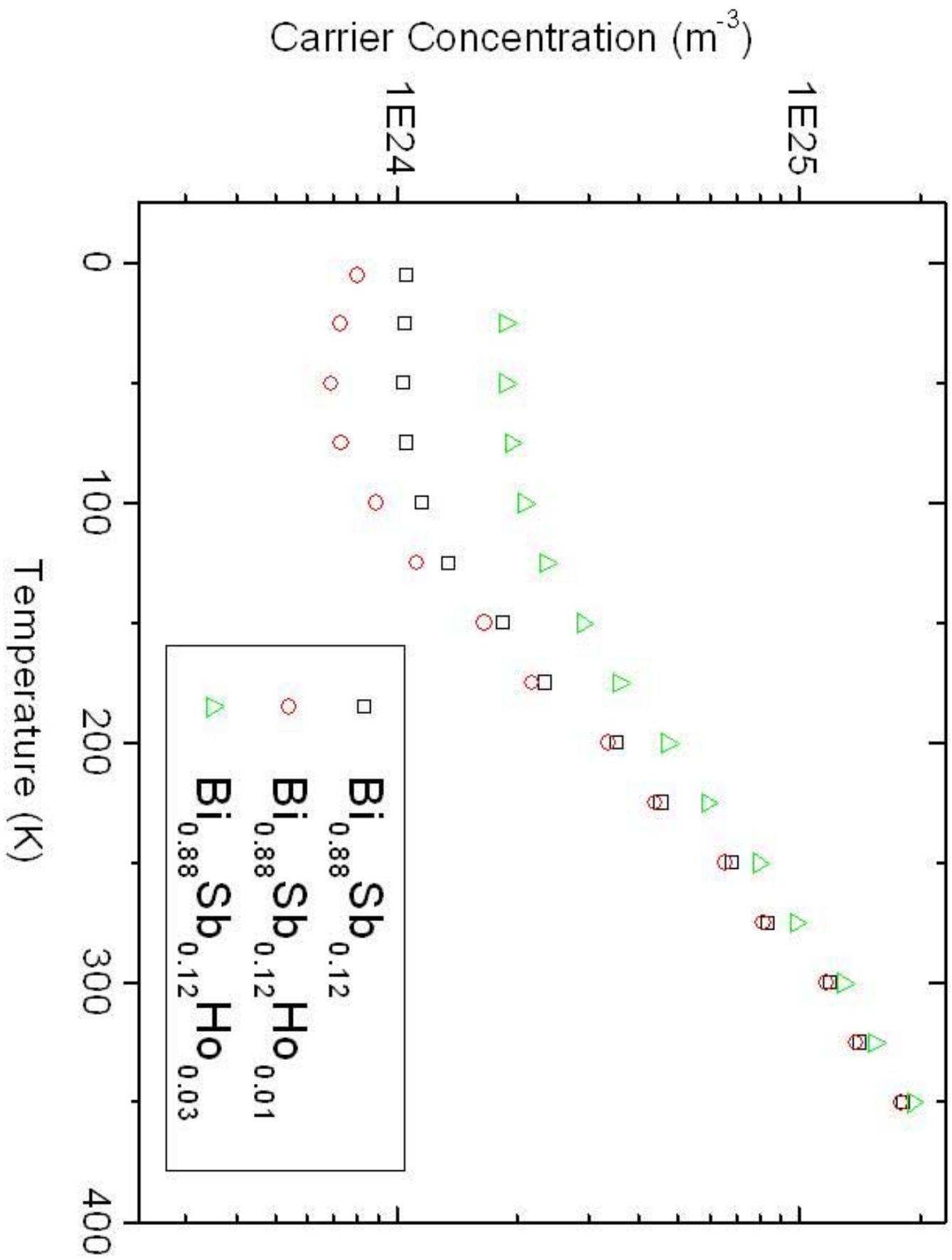

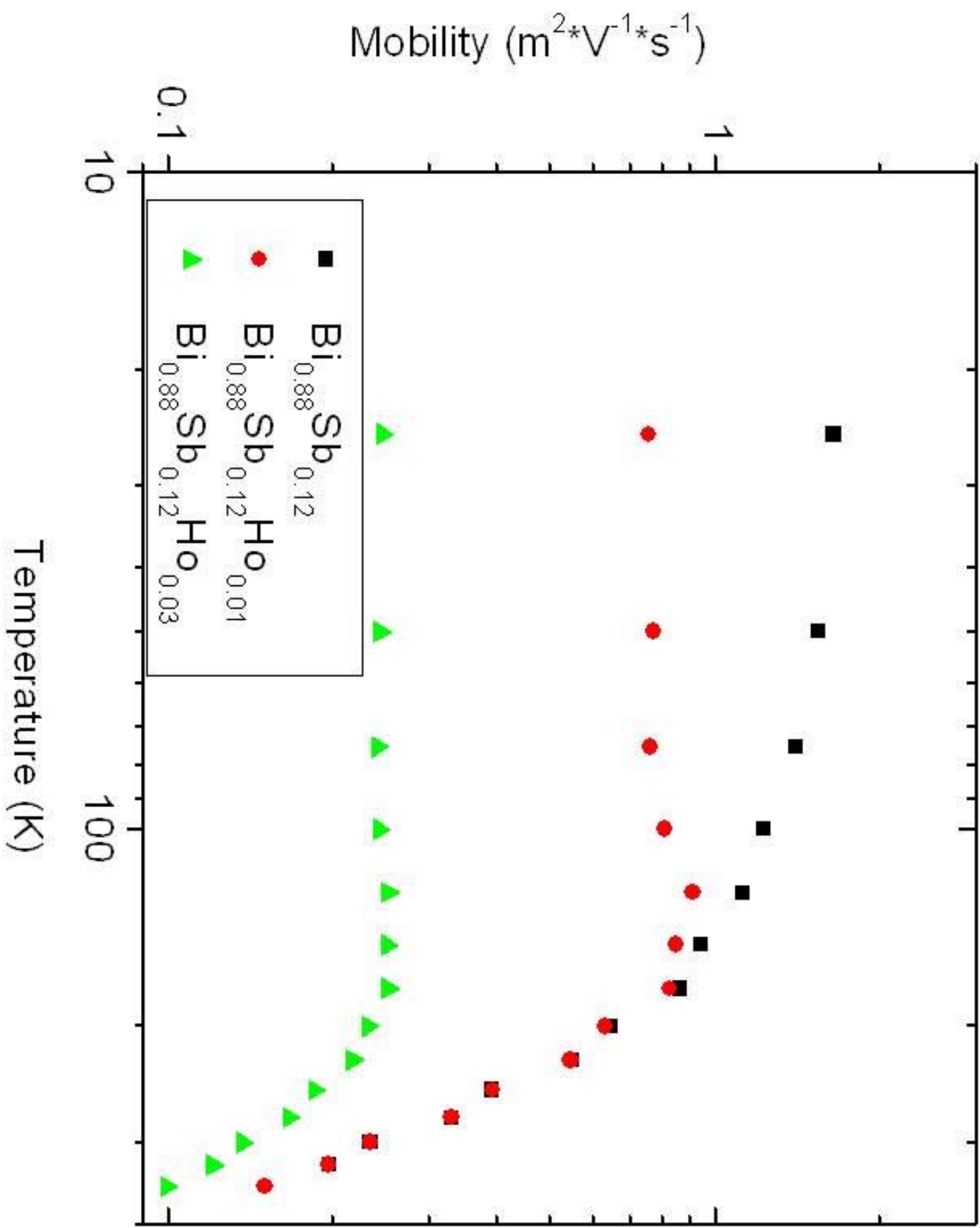

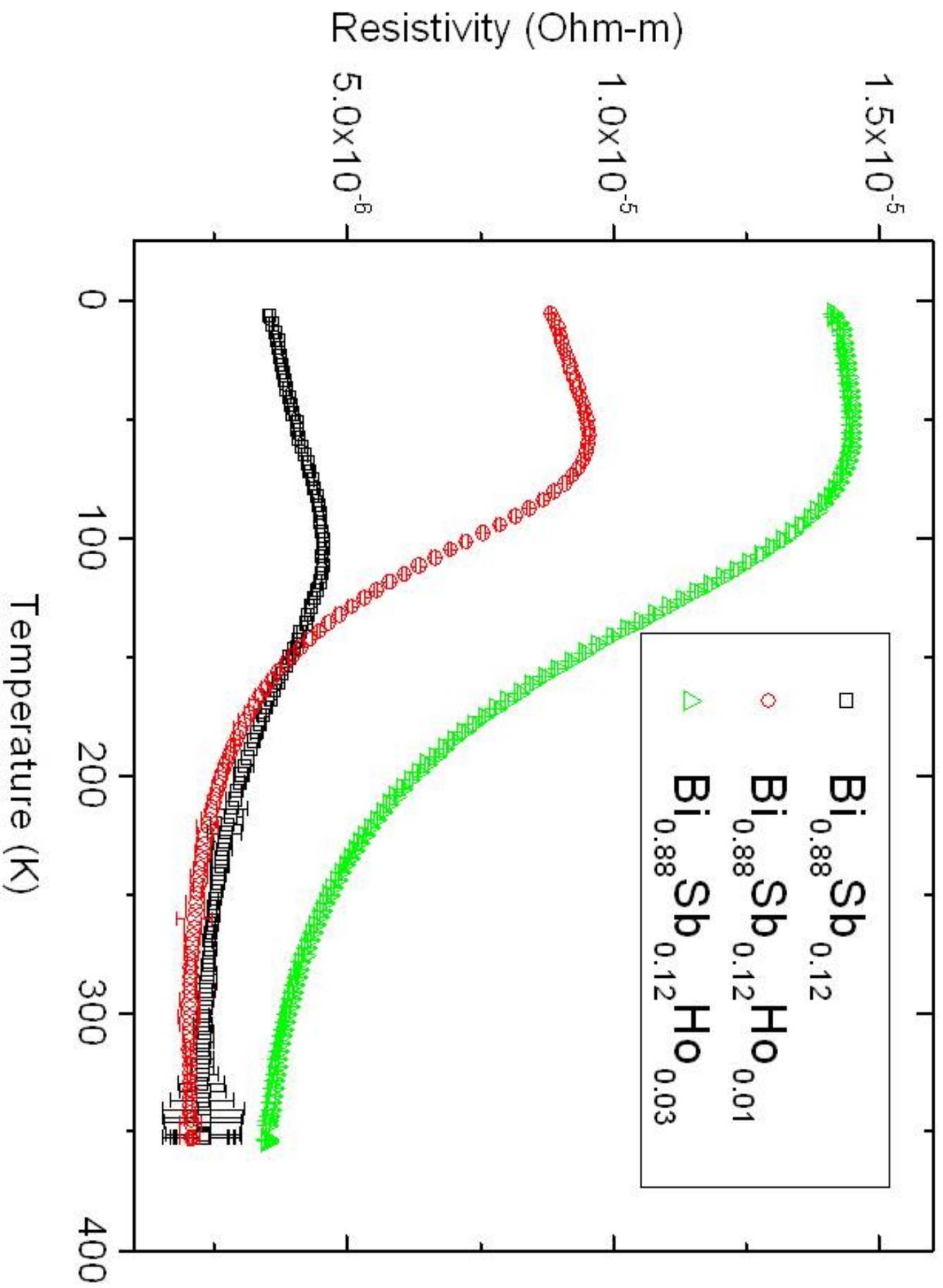

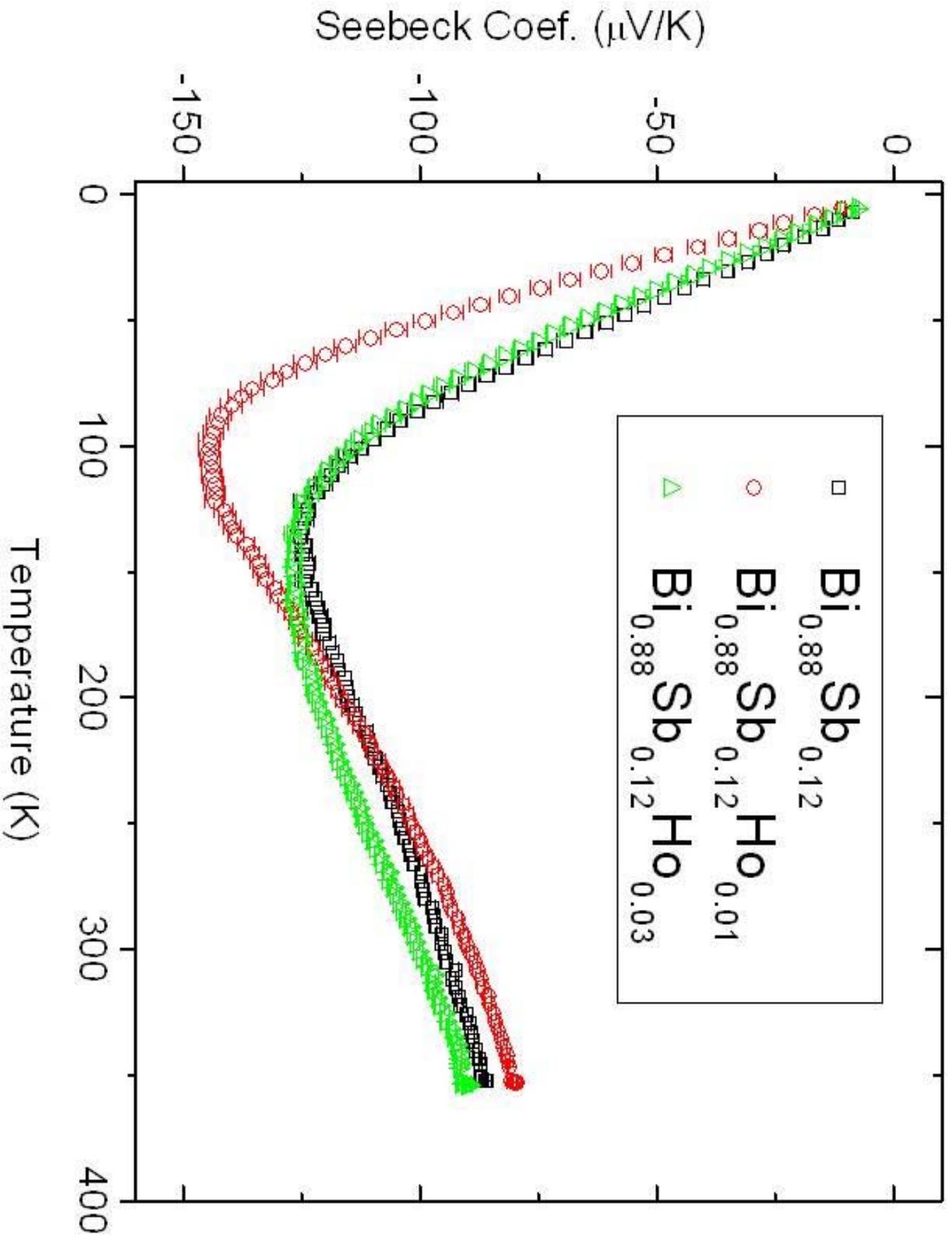

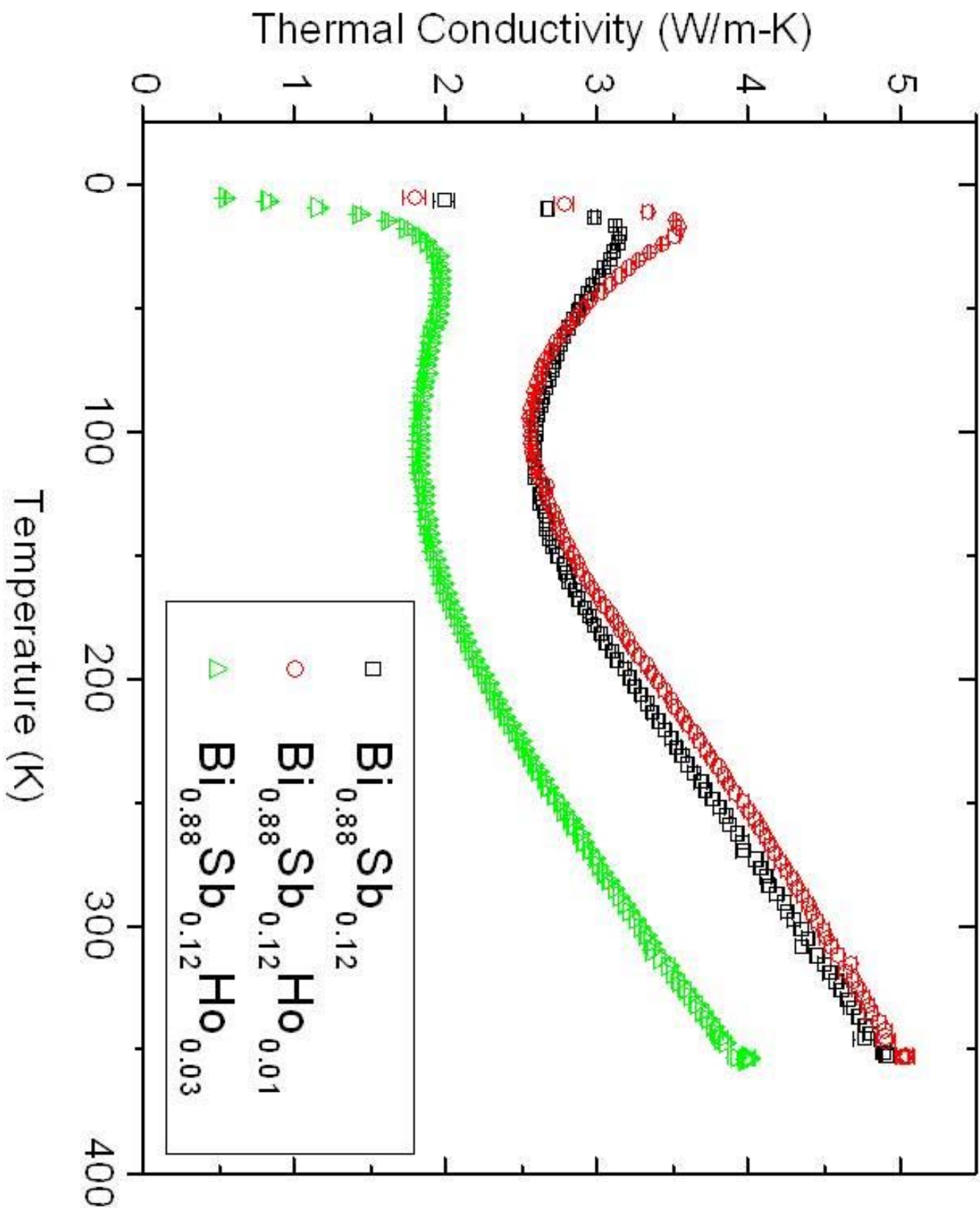

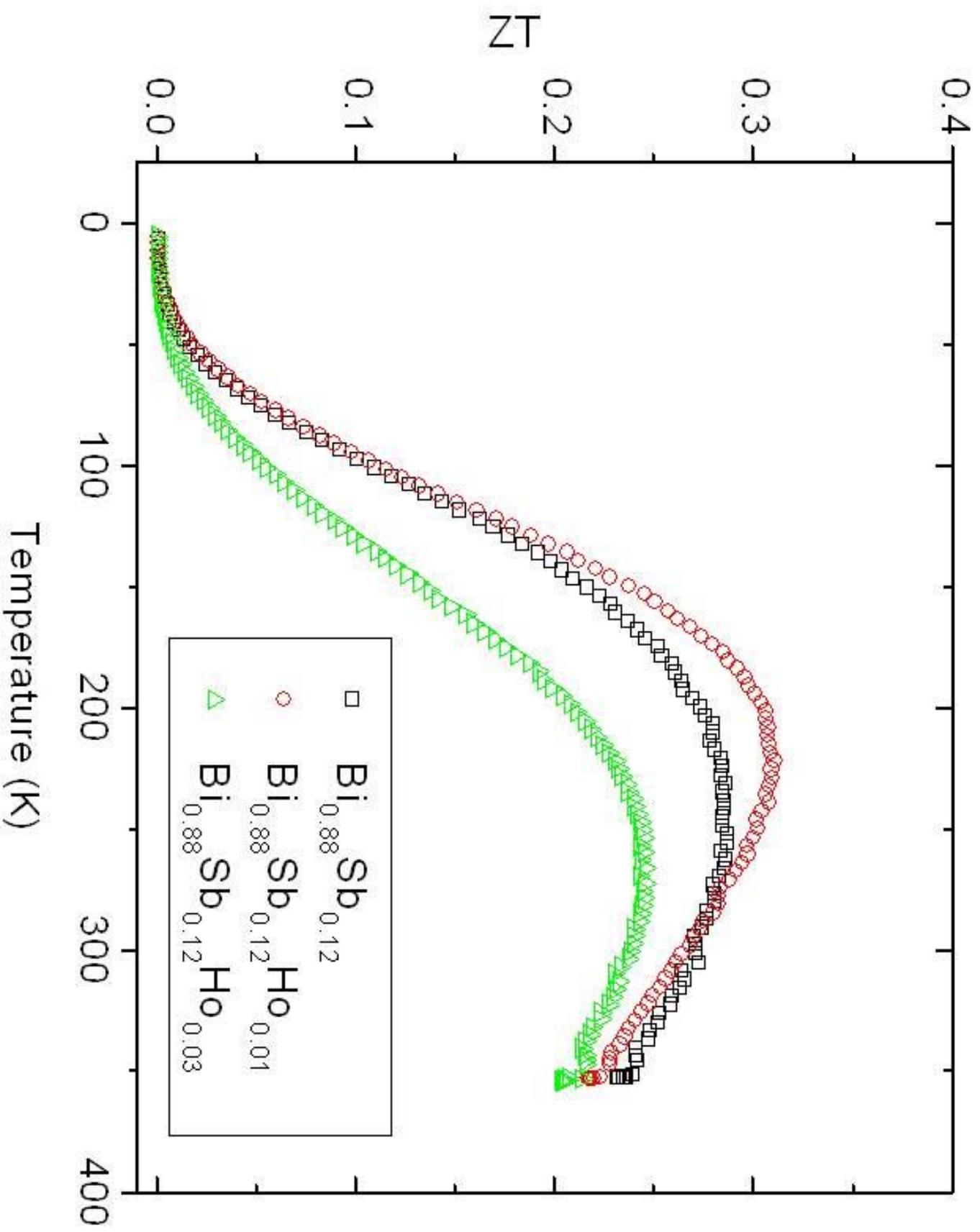